# FORMATIVE ASSESSMENT AND ITS E-LEARNING IMPLEMENTATION

SEIBU MARY JACOB[†]

School of Business, Swinburne University of Technology (Sarawak Campus)
Kuching, Sarawak, Malaysia
sjacob@swinburne.edu.my

BIJU ISSAC

School of IT & Multimedia, Swinburne University of Technology (Sarawak Campus)
Kuching, Sarawak, Malaysia
bissac@swinburne.edu.my

Innovation in assessment is no more a choice in a tech-savvy instant age. The purpose of this study was to get more insight into the implementation of formative assessment through the e-learning tool called Black Board (BB) Learning System in our University. The proposal is to implement a series of weekly or fortnightly tests on the BB. These would have options to provide sufficient feedback as a follow-up to the student's attempt. Responses from questionnaires were used to discover the important concerns in the students' perceptions on the proposed concept of continuous assessment and the BB Online test implementation. The results indicate that students support the idea mainly because they find the disintegration into small topic assessments as useful, coupled with the availability of immediate teacher feedback. What we intend is a culture of success, backed by a belief that all pupils can achieve the same.

Keywords: Formative assessment; Black board;

## 1. Introduction

We live in an era when the dynamic role of the teacher is emphasized to channel the personal, emotional and social pressures of youngsters to become better learners in future. Teachers can build in many opportunities to assess how students are learning and, accordingly make beneficial changes in instruction. This diagnostic use of assessment to provide feedback to teachers and students over the course of instruction is called formative assessment. It stands in contrast to summative assessment, which generally takes place after a period of instruction and requires making a judgment about the learning that has occurred (e.g., by grading or scoring a test or paper). This paper addresses the benefits of formative assessment and points to its workability through e-learning. Formative and summative approaches should not have to be mutually exclusive.

[†] Swinburne University of Technology (Sarawak Campus), Level 1, State Complex, Jalan Simpang Tiga, 93576 Kuching, Sarawak, Malaysia.

In theory, formative assessment should prepare students to excel on the summative year (semester) end tests.

## 2. Theoretical Support and Evidence of Formative Assessment

### 2.1. Student Excellence vs. Learning

Cross[1] details on three conditions to student excellence: high expectations, student participation & involvement, assessment and feedback. Quality assurance feedback points to the fact that the weakest link of the three is assessment & feedback. Assessment & feedback forms the core of learning and the student experience, according to Falchikov & Thomson[2]. The three conditions identified by Cross are not discrete items, but rather integrated and interdependent. It's an open secret that student involvement with assessment, whether peer or self-assessment, is still rare in higher education. Examining assessment practice is a useful means of gauging change and development in higher education, since it impacts directly or indirectly on other processes, as voiced by Hounsell[3].

### 2.2. Effectiveness of Formative Assessment

In order for Formative assessment to be effective: Teachers need to be proactive in the classroom (Vygotskian notion of scaffolding, 1978)[4], students need to be given thoughtful feedback on errors as well as on correct responses (Social Cognitive Theory, Bandura, 1986)[5], students then need to be given time to reflect on the feedback and their own performance (metacognition), students then work with the teacher to attain a more competent level of performance (scaffolding). From the reviews and Meta analyses of Formative assessment[6,7,8,9,10,11], the authors conclude that Formative assessment is an integral component of a productive classroom environment. Frequent testing should be combined with adequate feedback. Effectiveness also was shown, in a Meta analysis[12], to be dependent on the content of Formative assessment and its use within a structured program.

2.2.1. Formative assessment and feedback

Cross portrays learning without feedback metaphorically as learning archery in a darkened room. The focus here is on feedback as part of assessment and not on feedback as part of the classroom learning process. If the two components of tutor feedback and student learning are separated, the formative aspect of assessment is lost[13]. Any assessment can be beset by two negative outcomes[14]: an obsessive focus on competition, the attendant fear of failure on part of the low achievers. Quoting Lambert & Lines[15], the effectiveness of feedback can be maximized by conducting it at the level of the individual learners and avoid comparisons with other pupils, emphasizing individualized, challenging but achievable 'targets', keeping it criteria referenced with shared understanding between teacher and students and keeping precise recommendations for improvement of particular subject skills, concepts or knowledge. Three stages of Feedback can be identified as follows: recognition of the desired goal, evidence about the present position, and some understanding of the way to close the gap between the two. The four ways in which good formative assessment can influence learning are by[11] raising levels of motivation to learn, deciding what to learn, learning how to learn and evaluating learning.

## 3.  The E-Learning Implementation using Black Board Learning System

The e-learning implementation of formative assessment can be effectively done through Black Board learning system. Swinburne University of Technology (Sarawak Campus) uses Black Board Learning System (Release 6) to connect the students online and give access to the subject lecture slides and other learning materials. This is an excellent tool to help the students with the updated information on a subject through web technology. We explored the Online Test option on Black Board and found that to be a great way to do formative assessment as a series of 10 or so (as weekly or fortnightly) online assessments.  We created a Java Online Test on the Black board with 10 questions (6 multiple choice and 4 true or false type). Once the Java Online Test link is clicked on Black Board, it shows up with a screen to do the online test. The online assessment can be created using the options in the BB Test Canvas. The tests can be multiple choice, true/false, multiple answer, ordering, matching, fill in the blanks and essay type. Questions could be arranged in a random order, imported from an existing question pool or uploaded separately. As the students decide to take the test, a screen to preview the assessment and to select the right answer appears. Once the online test is taken, the review of assessment is presented to the students as follows in figure 1, with detailed feedbacks.  It may be noted that in the review of assessment, the students can see their score, and immediate feedbacks are given with detailed explanations. Some added options that can be thought of, when conducting such online assessments are as follows. The assessment can be made as an open book test to help the students feel a bit relaxed as they have to take such assessments on a weekly basis and the assessment questions can be made slightly indirect. Meaning, it doesn't give solutions to the students straight from the lecture notes. Some of the advantages of the e-learning implementation of formative assessment can be given as follows. It gives immediate feedback to the students so that the learning process happens without delay compared to traditional class room based approach, the options to create detailed feedbacks help the student to get to the root of his mistake, with sufficient explanation, it makes the learning aspect quite attractive to the students as they have to take the assessment online and it shows the scores to the students so that proper evaluation on one's standing in terms of subject knowledge is made clear. To enhance the feedback or online correspondence expected from the part of the lecturer, other Black Board tools like Discussion Forum and Virtual Class Room can be used.

Formative Assessment and its E-learning ImplementationFig.1.Review Assessment Screen. This shows the final score and the detailed and immediate feedback on the answers, explaining it to the students. The detailed feedback for the first question is shown.

## 4. Students' perceptions on Continuous assessment and the BB implementation

A survey was conducted among our University students, using a tested questionnaire consisting of 16 questions. The first part of the survey consisted of questions concerning the students' view of continuous assessment, necessity of teacher feedback. The reliability of this part, questioning whether the aspects were perceived by the students as intended by the designers of the assessment, was moderately high (coefficient alpha = 0.67). The second part was developed to measure their attitudes towards implementation on the Black Board System through the test option. The reliability of this part was high (coefficient alpha = 0.81).The third part of the survey consisted of open ended questions, to reflect on the advantages, disadvantages of the intended frequent assessments. A statistical analysis of the perceptions was done using SPSS. The difference among the Business, Engineering & IT streams, was statistically significant regarding the views on continuous assessment as open book assessments (Cramer's V=0.206, p<0.05). The Engineering students seemed to favor the option "Very Much Needed" in comparison to the Business and IT students as in figure 2.

4.1.1. Closed Questions

Frequent Internal Assessments – A majority of students were supportive of frequent internal assessments, as open book assessments. The top reasons were that learning happens in steps and not for final exams, it checks on learning, it provides constant feedback on the process. BB test vs. Classroom Test Option – They were also favouring the BB test option over the classroom test option, the major reasons being immediate feedback and flexibility/convenience. The students were in favour of splitting the course

261

Formative Assessment and its E-learning Implementation

work component into 10 or 12 weekly BB assessments, the major reasons being the easiness to study for small topics and not losing much marks even if one assessment was badly done. Teacher Feedback – The need for teacher feedback was viewed highly in 2 aspects namely, the feedback is necessary to measure my strengths and weaknesses and that the feedback helps me to check whether I am on track. Rating of BB test option – There was observed a difference in rating of the BB test option across the Business, Engineering & IT streams. The difference was statistically significant (Cramer's V=0.236, p<0.05).

4.1.2. Open Questions

Students found that the feedback was immediate and this could help them to correct their mistakes without delay. Also they thought it made sure they learn the individual topics progressively. They felt this aspect would prove good also for the summative semester examination. They agreed that the proposed method would help them to keep in touch with the subject throughout the semester. Students voiced their concerns on the learning outcomes (grades) fearing that "learning was overemphasized" in frequent assessments. Too many tests were thought of as bringing in stress/pressure to perform. To the question "Which one is better from a future career point of view-Good learning & more marks OR Less learning and more marks", there was a 60% response in favour of the former.

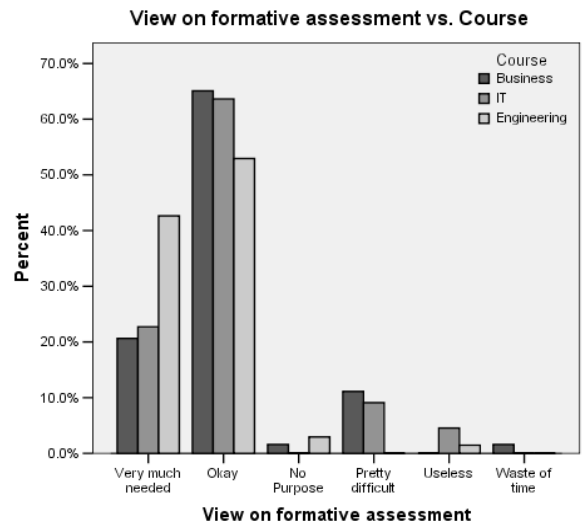

Fig.2. A general majority views formative assessment as okay. But the Engineering students voice it as 'very much needed' compared to the Business and IT students.

5. Conclusions

The ultimate user of assessment tasks to improve learning is the student. When the educational system focuses on rewards and ranking, then students are on a look out to obtain the best marks rather than to improve learning. 'The fear of failure' dominates and they avoid difficult tasks, looking only for clues to the right answer. These negative outcomes are not inevitable. 'While formative assessment can help all students, it yields particularly good results with low achievers by concentrating on specific problems with their work and giving them a clear understanding of what is wrong and how to put it right[14]'. Good formative assessment is not easy to achieve, taking into account the pressure from the public/parents, students themselves to produce results, and requires a leap of faith by the teaching community. The BB option is just one of the good effects of our technology driven times. The assessment to monitor & improve progress could be proved to improve summative grades.

**References**

1. Cross K.P., (1996) Improving Teaching and learning through classroom assessment and classroom research, in G. Gibbs (ed.) Improving Student Learning: using research to improve student learning (Oxford, Oxford Centre for Staff Development), pp. 3-10.
2. Falchikov, N. & Thomson, K. (1996), Approaches to studying and undergraduate well being, in G. Gibbs (ed.) Improving Student Learning: using research to improve student learning (Oxford, Oxford Centre for Staff Development).
3. Hounsell, D., McCulloch, M. & Scott, M. (1996), The ASSHE Inventory: changing assessment practices inn Scottish higher education (Centre for Teaching, Learning and Assessment, University of Edinburgh and Napier University).
4. Vygotsky,L.S. (1978).Mind and Society: The development of higher mental processes.(Cambridge, MA: Harvard University Press).
5. Bandura, A. (1986). Social Foundations of thought and action: A social cognitive theory. (Englewood Cliffs, NJ: Prentice-Hall).
6. Bangert Downs, R.L., Kulik,C.C., Kulik, J.A. & Morgan, M.T. (1991). The Instructional Effects of feedback in test-like events.
7. Black, P. & William, D. (1998), Assessment and Classroom learning. (Assessment in Education, 5 (1), 7-74.)
8. Crooks, T.J. (1988), The impact of classroom evaluation practices in students. (Review of Educational Research, 58, 438-481).
9. Kluger, A.N. & DeNisi, A. (1996), The effects of feedback interventions on performance: A historical review, a meta analysis, and a preliminary feedback intervention theory, (Psychological Bulletin, 119, 254-284).
10. Kulik, C.C., Kulik, J.A. & Bangert-Drowns, R.L. (1990), Effectiveness of mastery learning programs: A meta analysis, (Review of Educational Research, 60, 265-299).
11. Natriello, G. (1987), The impact of evaluation processes on students, (Educational Psychologist, 22, 155-175)
12. Fuchs, L.S. & Fuchs, D. (1986).Curriculum–based assessment of progress toward long-term and short-term goals. The Journal of Special Education, 20, 69-82.
13. Orsmond, P., Merry, S. & Reiling, K. The use of student derived marking criteria in peer and self assessment, Assessment & Evaluation in Higher Education, 25(1), (2000), pp21-38
14. Black, P. & William,D. Inside the Black Box: Raising Standards Through Classroom Assessment, Phi Delta Kappan, 80(2), (1998)
15. David Lambert and David Lines, Understanding Assessment-Purposes, Perceptions, Practice (Routledge Falmer, 2000)